\documentclass[showpacs,twocolumn,preprintnumbers,prb,amssymb]{revtex4}
\usepackage{graphicx}
\usepackage{bm}
\usepackage{amsmath}
\usepackage{amsfonts}
 
\newcommand{\la}{\langle}
\newcommand{\ra}{\rangle}
\DeclareMathOperator{\imag}{Im}
\DeclareMathOperator{\real}{Re}                

\def \ve{\varepsilon}
\def \si{\sigma}
\def \w{\omega}
\def \mu{\zeta}

\begin{document}
\title{Efficient Linear Scaling Approach for Computing the Kubo Hall Conductivity}
\author{Frank Ortmann$^{1}$}
\author{Nicolas Leconte$^{2}$}
\author{Stephan Roche$^{2,3}$}
\affiliation{$^1$ Institute for Materials Science and Dresden Center for Computational Materials Science, Technische Universit\"{a}t Dresden, 01062 Dresden, Germany}
\affiliation{$^2$ICN2—Institut Catala de Nanociencia i Nanotecnologia, Campus UAB, 08193 Bellaterra (Barcelona), Spain}
\affiliation{$^3$ICREA, Instituci\'{o} Catalana de Recerca i Estudis Avan\c{c}ats, 08070 Barcelona, Spain }

\begin{abstract}
We report an order-{\it N } approach to compute the Kubo Hall conductivity for disorderd two-dimensional systems reaching tens of millions of orbitals, and realistic values of the applied external magnetic fields (as low as a few Tesla). A time-evolution scheme is employed to evaluate the Hall conductivity $\sigma_{xy}$ using a wavepacket propagation method and a continued fraction expansion for the computation of diagonal and off-diagonal  matrix elements of the Green functions. The validity of the method is demonstrated by comparison of results with brute-force diagonalization of the Kubo formula, using (disordered) graphene as system of study. This approach to mesoscopic system sizes is opening an unprecedented perspective for so-called  {\it reverse engineering} in which the available experimental transport data are used to get a deeper understanding of the microscopic structure of the samples. Besides, this will not only allow addressing subtle issues in terms of resistance standardization of large scale materials (such as wafer scale polycrystalline graphene), but will also enable the discovery of new quantum transport phenomena in complex two-dimensional materials, out of reach with classical methods.
\end{abstract}

\maketitle

\section{Introduction} 
The Hall effect is one of the central phenomenon in Condensed Matter physics with great importance for measuring carrier density and mobility in semiconductors. The long history started almost 150 years ago when E. H. Hall observed a transverse voltage drop over a conducting bar driven by a longitudinal current \cite{Hall_original} and it experienced an important turn in 1980 when K. von Klitzing observed a quantized version of this effect, the so-called Quantum Hall Effect (QHE)\cite{vonKlitzing_original}.
This had an immediate impact and triggered huge amount of work which led to the definition of a resistance standard.\cite{Porier:2010:NAT}

The effect of disorder and related localization phenomenon is central to understand the integer QHE from a bulk perspective, where $\sigma_{xy}$ plateaus develop by varying the charge density, while the plateau  region coincide with a region of vanishing $\sigma_{xx}$. \cite{Aoki87} This is explained by Anderson localization of the wave functions in the Landau levels induced by disorder. The most standard method to treat bulk conductivities microscopically is the linear-response theory, or
the Kubo formula, which was first applied by Aoki and Ando \cite{AokiSSC81} to the QHE problem \cite{Aoki87}. Such quantity can be connected to measurements in the Corbino geometry for which electrodes are attached to the inner and outer perimeters of an annular sample, which allow to investigate bulk transport physics in the high magnetic field regime.

Recently, the QHE has played a crucial role for the first unambigous proof of the existence of single atomic layers of graphene \cite{Graphene:NAT2005,Graphene_Kim:2005,Novoselov:2007:RoomtemperatureQHE}. Indeed, in graphene, the peculiar Berry phase related to the pseudospin quantum degree of freedom results in a unique spectrum with fourfold degenerate Landau levels  (due to the spin and valley degeneracies), and half-integer QHE with Hall plateaus emerging at $\sigma_{xy}=\frac{4e^{2}}{h}(n+1/2)$, with integers $n$. QHE related research in graphene is a vivid field, studying electron-electron interaction \cite{Nomura:2006,Young-Kim:2012}, superlattices \cite{Ponomarenko:2013:Superlattice,Dean:2013:Hofstadter,Hunt:2013:Hofstadter,Yang:2013:GBN}, graphene functionalization and topological currents.
Additional transverse transport phenomena are being studied such as (quantum-) anomaleous Hall effect,\cite{Nagaosa:2010:RMP,QAHE:2013:SCI}, 
spin Hall effect \cite{Kato:2004:SHE} and quantum spin Hall effect. \cite{Konig:QSH:2007}

The role of disorder in graphene is also debated and the QHE features, with very robust behavior to large disorder \cite{Guillemette:PRL:2013}, also present additional peculiarities depending on the symmetry breaking aspects conveyed by defects.\cite{Ostrovsky:2006:PRB,Tworzydlo:2006:PRL,Morpurgo:2006:PRL,Peres:2006:PRB,Gattenlohner:PRL:2014} Recently, we have studied the case of oxidized graphene, and have shown the formation of disorder-induced resonant critical states appearing in the zero-energy Landau level with finite $\sigma_{xx}$ and suggesting a zero-energy $\sigma_{xy}$ quantized plateau\cite{Leconte:2014:QHEEpoxy}. The confirmation of such a plateau will demand to revise the description of topological invariant in disordered graphene, an exciting direction of work. However a real space calculation of $\sigma_{xy}$ is needed and demand for the development of new methodology.

Indeed, the importance of an efficient computational methodology for the Hall conductivity $\sigma_{xy}$ stands in sharp contrast to the limited possibilities to simulate this quantity with usual approaches. One of them starts with the Kubo formula and requires the knowledge of the full eigenstates of a given system. \cite{AokiSSC81}. Such formulation allows for advanced analysis\cite{Sheng2006,Goswami:2008:PRB} but suffers from unfavourable cubic scaling with system 
size which restricts to small systems because diagonalization is necessary. All eigenstates are needed and have to be combined to calculate $\sigma_{xy}$. This computational limitation makes the analysis of realistic models of disordered samples \textit{a priori} impossible. Additionally, such approach is restricted to unreasonably strong magnetic fields that exceed by far what is achievable experimentally, and forces the magnetic length scale to dominate over all other length scales of the problem (mean free path, average distance between impurities,...) limiting the exploration of complex magnetotransport in intermediate 
or even low-magnetic field regimes, thus calling for approaches with improved numerical performance.\cite{Ortmann:2013:QHE,Rappoport:2014ARX,Ishii:PRB2011}

In this paper, we present a highly efficient order-$N$ algorithm for the computation of the Kubo Hall conductivity that allows us to circumvent aforementioned limitations, and offer fascinating perspectives for exploring transverse transport phenomena in disordered two-dimensional materials with almost arbitrary complexity. A validation of the method on several models of clean or disordered graphene is made by comparing this new time evolution method with brute force diagonalization technique. Some preliminary illustration of the method have been published in a prior work \cite{Ortmann:2013:QHE}. The method is inspired by the real space implementation of the dissipative Hall conductivity at zero frequency, which has proven undisputed computational efficiency and predictive power (see Ref.\cite{RocheMayou:1997,FOA_2014} for computational details, and Ref. \cite{ROC_SSC152} for some illustration on realistic models of disordered graphene).

\section{Methodology}
{\it Kubo formalism -- }
The general framework is provided by Kubo's linear response theory\cite{KuboOrig,Mahan} 
in which the dc conductivity is given as 

\begin{equation}\label{KubosConductivity}
\si_{xy}(\w=0)=\frac{1}{V}\int_{0}^{\infty}dt \int_0^\beta d\lambda
\text{Tr}\left[\rho_0 j_y j_{x}(t+i\hbar\lambda) \right]
\end{equation}
Starting from Eq. \eqref{KubosConductivity}, the evaluation of the conductivity $\sigma_{xy}$
proceeds by introducing an orthonormal many-particle basis 
with $\la n|H|n\ra=E_n$ which allows to rewrite Eq. \eqref{KubosConductivity} into
\begin{equation}
\si_{xy}=-\frac{2}{V}\int_{0}^{\infty}dt \\
\lim_{\eta\to 0^+}\sum_n
\real\la n|\rho_0 j_y \frac{1}{E_n-H+i\eta} j_x(t) |n\ra
\end{equation}
where $\eta$ is a small parameter necessary to converge this expression.

In the case of non-interacting electrons one can proceed further 
towards the single-particle picture and obtain
\begin{equation}
\begin{aligned}
\si_{xy}&=-\frac{2}{V}\int_{0}^{\infty}dt  \lim_{\eta\to 0^+}\sum_{k} f(\ve_{k}-\mu)\\
&\real\left[\la k| j_{y}\frac{1}{\ve_{k}-H+i\eta}j_{x}(t) |k\ra \right]
\end{aligned}
\end{equation}
where the Fermi-Dirac distribution is described by $f(\ve_{k}-\mu)$ and the chemical potential $\mu$ has been introduced. The single-particle eigenvalues $\ve_{k}$ can in principle be obtained by matrix diagonalization of finite systems which has been widely employed in the literature for studying model systems with phenomenological description of disorder. Unfortunately such numerical operations quickly makes the methodology computationally prohibitive, in particular for simulating complex and large system sizes closer to the experimental ones and at moderate magnetic fields.
Clearly, the computational problem then becomes untreatable with diagonalization-based methods, 
i.e. it becomes impossible to calculate (and store) all eigenvalues and eigenvectors, so that time-evolution based
approaches appear promising. 
By introducing $\int_{-\infty}^{\infty}dE \delta(E-H)$ we arrive at an alternative representation of the Hall conductivity
\begin{equation}\label{Cond_noEV}
\begin{aligned}
\si_{xy}&=-\frac{2N_s}{V}
\int_{0}^{\infty}dt 
\int_{-\infty}^{\infty}dE f(E-\mu)\\
&\lim_{\eta\to 0^+} 
\real\left[\la \Psi_1|\delta(E-H) j_{y}\frac{1}{E-H+i\eta}j_{x}(t) |\Psi_1\ra \right]
\end{aligned}
\end{equation}
which proves useful for the numerical implementation. 
In Eq. \eqref{Cond_noEV}, the trace is replaced by an average over a random-phase state 
$|\Psi_1\ra$ according to $\sum_k\la k|...|k\ra\to N_s\la\Psi_1|...|\Psi_1\ra$. Such states are normalized by using the number of sites $N_s$.
We further introduce the projection operator 
$\sum_{j=1}^{N_R}|\Psi_j\ra\la\Psi_j|$, as well as the time-dependent
 $\si(t)$ fulfilling $\si_{xy}=\lim_{t\to \infty}\si_{xy}(t)$ with
\begin{equation}\label{num1}
\begin{aligned}
\si_{xy}(t')&=\frac{2}{\pi V_s}
\int_{0}^{t'}dt 
\int_{-\infty}^{\infty}dE f(E-\mu)\\
&\lim_{\eta\to 0^+}
\sum_{j=1}^{N_R}\imag\left[\kappa_{j}(E)\right]\\
&\real\left[\la\Psi_j| j_{y}U^\dag(t)\frac{1}{E-H+i\eta}
j_{x} U(t) |\Psi_1\ra \right],
\end{aligned}
\end{equation}
where 
\begin{equation}\label{kappa}
\kappa_{j}(E)=\la\Psi_j|\frac{1}{z-H}| \Psi_1 \ra
\end{equation}
are the the matrix elements of the Green function (with $z=E+i\eta$) and $V_s=V/N_s$ is the volume per site.
Equation \eqref{num1} is the basis for the numerical evaluation of the Hall conductivity  for which one eventually takes the limit $t'\to \infty$. One notes, however, that the result does not converge in all cases when taking this limit and an oscillatory behaviour of the integral is generally observed, particularly for the case of clean systems in high magnetic fields (and well separate Landau levels), which is the usual test reference for Kubo Hall transport simulations.

The reason is that the Kubo formula Eq. \eqref{KubosConductivity}
assumes that at ``infinite past'' times, the system is in equilibrium, i.e.
at zero electric field, the current should vanish. In other words,
the current-current correlation function should decay with time difference $t$.
This should be fulfilled also for the final result.
In pristine systems this is not the case and must be imposed {\it a posteriori}.
In order to take this into account an additional small parameter $\delta>0$ 
has to be introduced such that the integral 
\begin{equation}\label{final}
\begin{aligned}
\si_{xy}(t')&=\frac{2}{\pi V_s}
\int_{0}^{t'}dt 
\int_{-\infty}^{\infty}dE f(E-\mu)\\
&\lim_{\eta\to 0^+}
\sum_{j=1}^{N_R}\imag\left[\kappa_{j}(E)\right]\\
&\real\left[\la\Psi_j| j_{y}U^\dag(t)\frac{1}{E-H+i\eta}
j_{x} U(t) |\Psi_1\ra \right]e^{-\delta t'}. 
\end{aligned}
\end{equation}
converges. 

We note that the same parameter has to be included in full analogy in the standard expression
for the calculation of the Hall conductivity based upon the knowledge of the eigensystem. We here display such formula \cite{AokiSSC81,AokiPRL85}
\begin{equation}\label{AokiFormula}
\begin{aligned}
\si_{xy}(t')&=\frac{\hbar}{V}
\lim_{\eta\to 0^+}\lim_{\delta\to 0^+}\sum_{k_1k_2}
f(\ve_{k_1}-\mu)\\
&\left[\frac{\la k_1|j_x|k_2\ra \la k_2|j_y|k_1\ra}{(\ve_{k_1}-\ve_{k_2}+i\eta)(\ve_{k_1}-\ve_{k_2}-i\delta)}-\text{c.c.}\right]
\end{aligned}
\end{equation}
which is identical to Ref. \cite{AokiPRL85} for the reasonable choice that $\eta=\delta$.
These two equations \eqref{final} and \eqref{AokiFormula} are comparable and will be evaluated for a couple of examples in Sect. \ref{Sec:Results} as numerical checks.

{\it Numerical implementation -- }
The calculation proceeds by splitting the time-dependent and time independent parts ($\kappa_j(E)$) while the sum over $j$ in \eqref{final}
controls the energy dependent convergence (see details below).
Firstly, in order to evaluate Eq. \eqref{kappa}, we make use of the Lanczos tridiagonalization of the Hamiltonian $H$
which allows us to use the continued fraction expansion for the diagonal Green function. From the diagonal elements
one can obtain the necessary off-diagonal elements in a second iterative step. We find a simple recursion relation connecting both which reads for $n>1$
\begin{equation}
\kappa_{n+1}=\frac{1}{b_n}\left[-b_{n-1}\kappa_{n-1}-(z-a_n)\kappa_n\right]
\end{equation}
while the first step in the iteration is given as $\kappa_2=\frac{1}{b_1}\left[1-(z-a_1)\kappa_1\right]$.

Secondly, the time evolutions of the quantities $|\Psi_1\ra$ and $|j_y\Psi_j\ra$
can be performed efficiently through a Chebychev expansion of the time-evolution operator $U(t)$ which
usually converges quite rapidly. For the final calculation of the brackets in Eq. \eqref{final} we
have to calculate off diagonal matrix elements with the advanced Green operator $(z-H)^{-1}$.
For the calculation of this second set of off-diagonal Green functions in Eq.
\eqref{final}, we use the continued fraction expansion. 
All operations are linear in sample size and
this holds also for the total expression since the scalar product in $j$ is restricted to a cutoff value $N_R$.
We use $\delta=0.002$ if not stated otherwise throughout the paper.

{\it System description}
The electronic properties of graphene can be described by the standard orthogonal tight-binding model
\begin{equation}
{\cal H} =  \sum_{\alpha}V_{\alpha}|\alpha \rangle\langle \alpha
|- \gamma_0 \sum_{\langle\alpha,\beta\rangle } e^{-i\varphi_{\alpha
\beta}} |\alpha\rangle\langle\beta|,
\label{eq:Hamiltonian}
\end{equation}
where $\gamma_0=2.7 eV$ is the nearest neighbor transfer integral and $V_{\alpha}$ the on-site energy, which can be chosen to describe various disorder models.
Spatially uncorrelated Anderson disorder is introduced through on-site energies taken at random from $[-W\gamma_{0}/2, W\gamma_{0}/2]$ where $W$ gives the disorder strength. This is a commonly used disorder model for exploring the metal-insulator transition in low-dimensional systems \cite{Evers2008,Sheng2006}, which, in average, does not break electron-hole symmetry or inversion symmetry. The inversion symmetry of the system related to the symmetry of AB sublattices can be artificially broken by a staggered potential of the form $V_{\alpha}=+(-)V_{AB}$ for A(B) sublattices (including all atoms). A less artificial variant of such potential can be introduced by diluting such sublattice terms for a weaker and random distribution, which mimicks some correlation in the potential beyond the Anderson model.

The magnetic field is implemented through a Peierls phase~\cite{Peierls} added to $\gamma_0$ which determines the magnetic flux to $\phi=\oint{\bf A}\cdot d{\bf l}=h/e\sum_{\rm hexagon}\varphi_{\alpha\beta}$ per plaquette. \cite{Ortmann:2011:EPL}
Spin degeneracy is assumed throughout the paper and an anti-symmetrization procedure for $\sigma_{xy}(E)$  has been performed consistent with electron-hole symmetry
for all considered disorder potentials.

\section{Results}\label{Sec:Results}

\begin{figure}[t]
\begin{center}
\includegraphics[width=0.8\columnwidth]{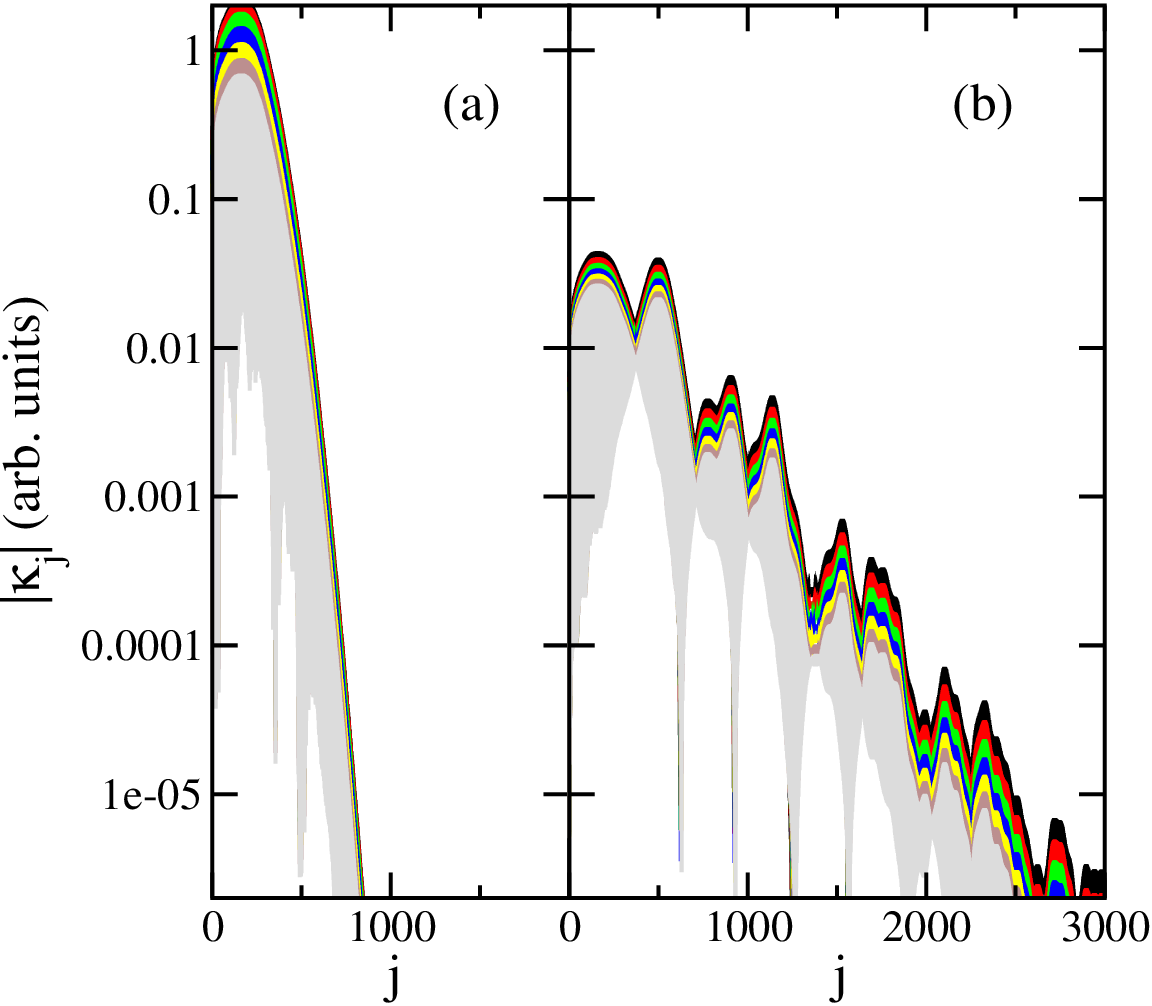}
\caption{(color online) Absolute values of the off-diagonal matrix elements Eq. \eqref{kappa} for pristine graphene (a) and disordered graphene (b). Different curves correspond to energies around the lowest-energy Landau level.}
\label{Fig0}
\end{center}
\end{figure}
Before we enter into the discussion of the physical results, we illustrate the convergence behaviour of the numerical algorithm which can be controled by the expression \eqref{kappa}.
A key quantity for the convergence is $\kappa_j(E)$ which can be studied prior to the time evolution of wavepackets and allows to estimate the required number of recursion steps which depend on the specific physical details (e.g. disorder, magnetic field), on the selected energy, and on the broadening $\eta$. Fig. \ref{Fig0} presents the modulus of $\kappa$ plotted versus $j$ for some selected energies in the case of pristine graphene (a) and disordered graphene (b). As a general trend, Fig. \ref{Fig0} shows that in case of pristine systems, the convergence of the Lanczos recursion can be reached with a low number of recursion steps ($N_R$), while increasing disorder requires $N_R$ to be increased typically to a few thousands.

\begin{figure}[t]
\begin{center}
\includegraphics[width=\columnwidth]{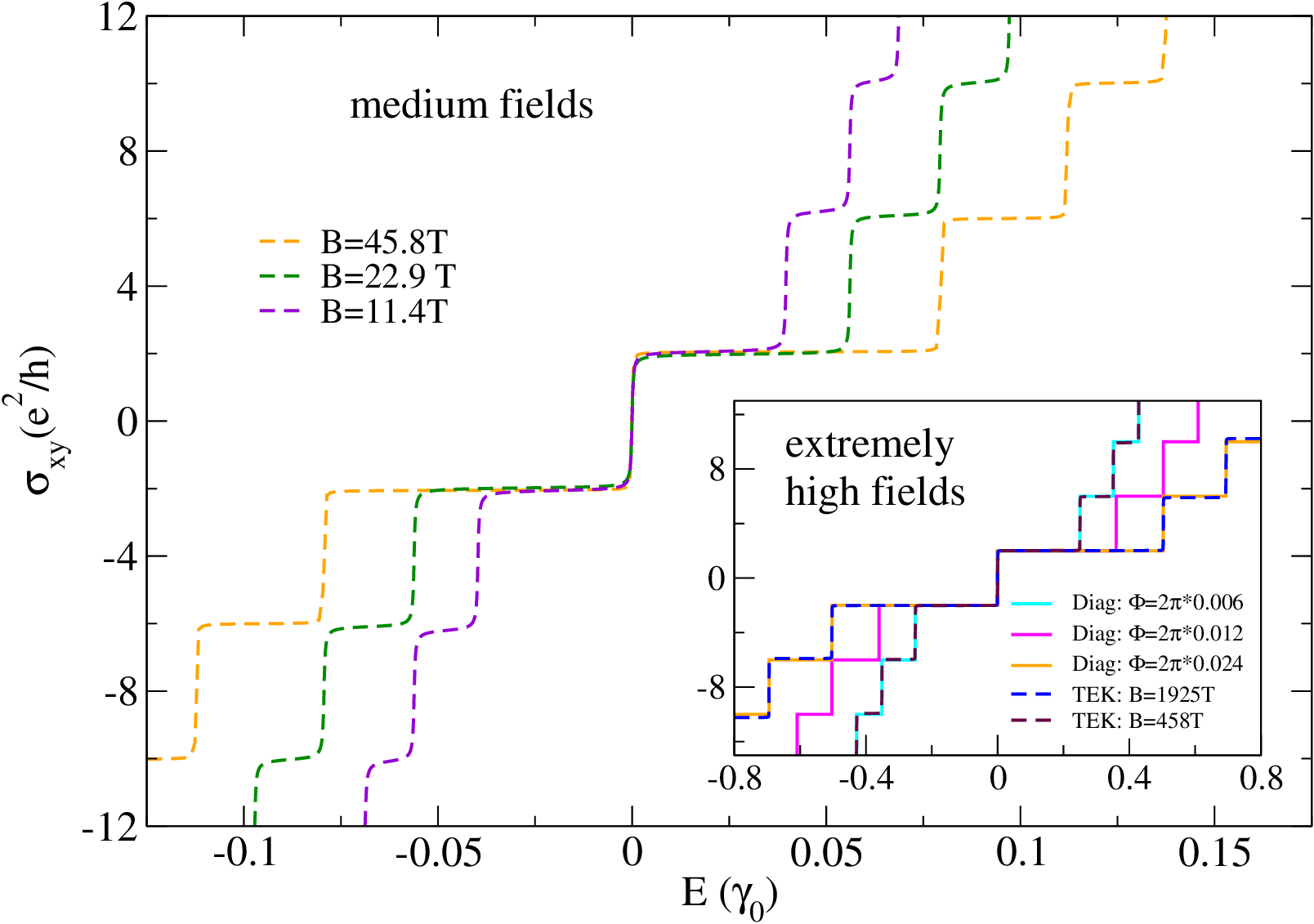}
\caption{(color online)  Hall conductivity for varying magnetic field strength. Main frame: moderately high fields; lower inset: extremely fields. 
Dashed curves in all frames indicate simulations with time-evolution Kubo (TEK)  approach. Note the magnetic-field scaling of Landau level energies $E\propto \sqrt{B}$ which is
reflected in varying energy scales of main frame and inset ($\gamma_0$ units).}
\label{Fig1}
\end{center}
\end{figure}

Turning to the simulations of the Hall conductivity, we first discuss the results for pristine graphene in Fig. \ref{Fig1} which shows the comparison of diagonalization method and time-evolution Kubo approach.
Figure 1 (inset) shows the case of extremely high magnetic fields (fluxes). We recall that, in order to fulfill the boundary conditions of the periodic system, the limit of very high fields is usually considered by diagonalization-based studies, because only small sample sizes can be treated by matrix diagonalization. The magnetic field is simultaneously given in the inset of Fig. \ref{Fig1} in terms of the total flux penetrating the sample. The corresponding results from the time-evolution Kubo method (TEK) is plotted as dashed lines for the same magnetic fields (values indicated in the inset legend). For the selected energies, the curves coincide visually. The square root scaling of their position with magnetic field is evident and Hall-conductance steps occur at the expected positions and with the expected height corresponding to the half-integer sequence $\sigma_{xy}=\pm (1/2+n)4e^2/h$. 

Next, we consider the interesting case of moderately high magnetic fields such as frequently observed in current experiments\cite{Young-Kim:2012,Ponomarenko:2013:Superlattice,Dean:2013:Hofstadter,Hunt:2013:Hofstadter,Yang:2013:GBN} in the main frame of Fig. \ref{Fig1}.
The Kubo approach allows to reduce magnetic fields down to only few Teslas where the first Landau levels appear at Fermi energies of few tens of meV.
Still the plateau sequence is obtained with very good accuracy, which shows that such approach can be of great practical use. 
Most notably, the figure demonstrates conductance quantization for fields as low as 11 Tesla. 
As an effect of finite energy resolution, we observe that the relative sharpness of the steps is reduced with lowering the field in this regime.
The conductance slope starts to become noticable as Landau levels get closer (as seen for 11.4 T at $\sigma_{xy}=\pm10e^2/h$). 
This consequence of the considered energy resolution (here $\eta=0.00025$) can be systematically improved further by increasing the number of recursion steps and decreasing $\eta$.

\begin{figure}[t]
\begin{center}
\includegraphics[width=\columnwidth]{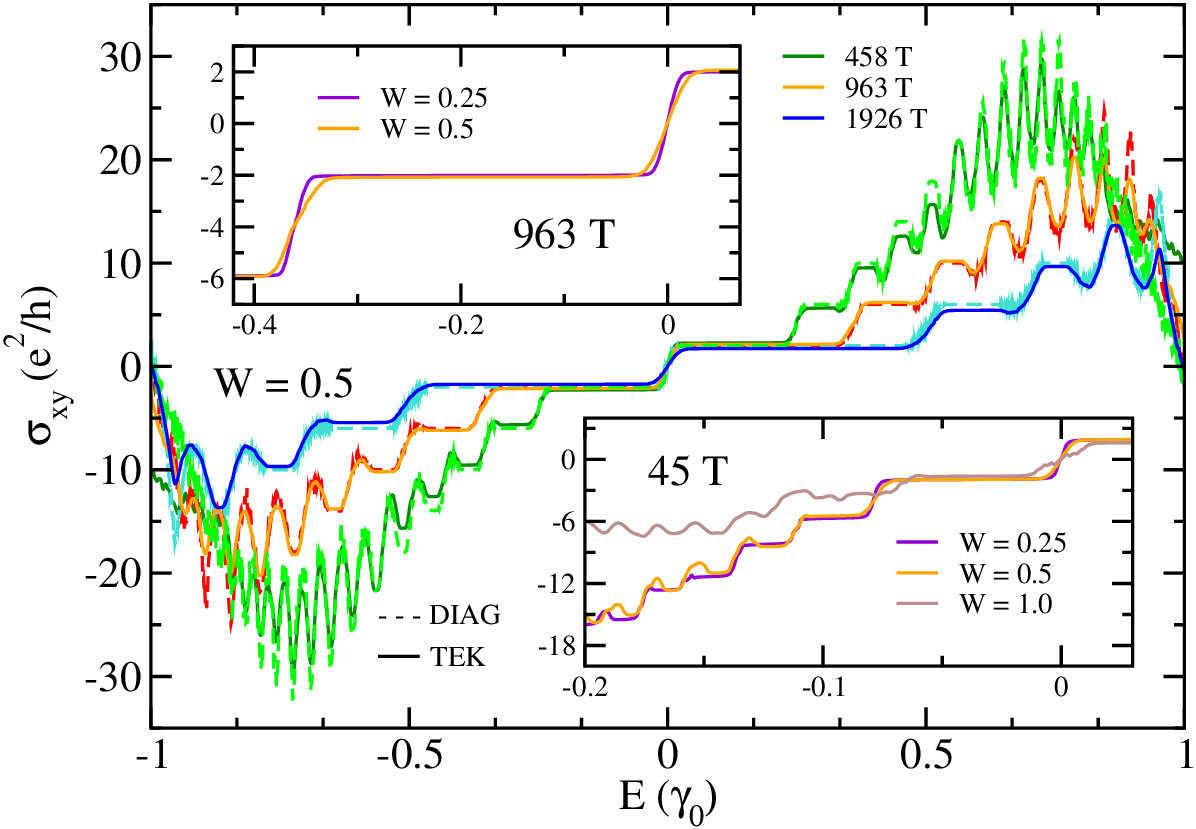}
\caption{(color online). Hall conductivity for Anderson disorder (dashed curves: exact diagonalization; solid curves: TEK) at different magnetic fields (main frame). Effect of ncreasing disorder for high field 963T (upper inset) and intermediate fields of 45T (lower inset). }
\label{Fig2}
\end{center}
\end{figure}

As a second example to illustrate the performance of the algorithm, we focus on the case of homogeneously disordered graphene. The case of Anderson disorder is chosen for Fig.~\ref{Fig2}, because this disorder is expected to induce a transition from the QHE regime to the conventional Anderson insulating state for sufficiently high value of $W$. The main-frame indicates very good agreement (within few percents) obtained between the Kubo algorithm and exact diagonalization techniques at low energy for very high magnetic fields. At higher energy, $\eta$ allows for fast convergence, at the expense of small loss in information compared to the exact method when Landau levels are getting closer in energy. To illustrate the effect of increasing disorder, a zoom on the first two Hall steps is provided for $963$ T in the upper inset. The Landau levels are broadened with increasing $W$ around the critical states, as expected from perturbation theory. Simultaneously, the first and second plateaus remain at $\pm 2 e^2/h$ and $\pm 6 e^2/h$ indicating robust QHE. In the lower inset, a more realistic magnetic field is chosen ($45$ T) to illustrate (i) the performance of the algorithm to probe low magnetic fields and (ii) the possibility to destroy conductance quantization at higher energy for $W=1$. For such disorder, only the zero-energy Landau level fully develops, while all states become localized for $E>E_1$.

\begin{figure}[t]
\begin{center}
\includegraphics[width=0.9\columnwidth]{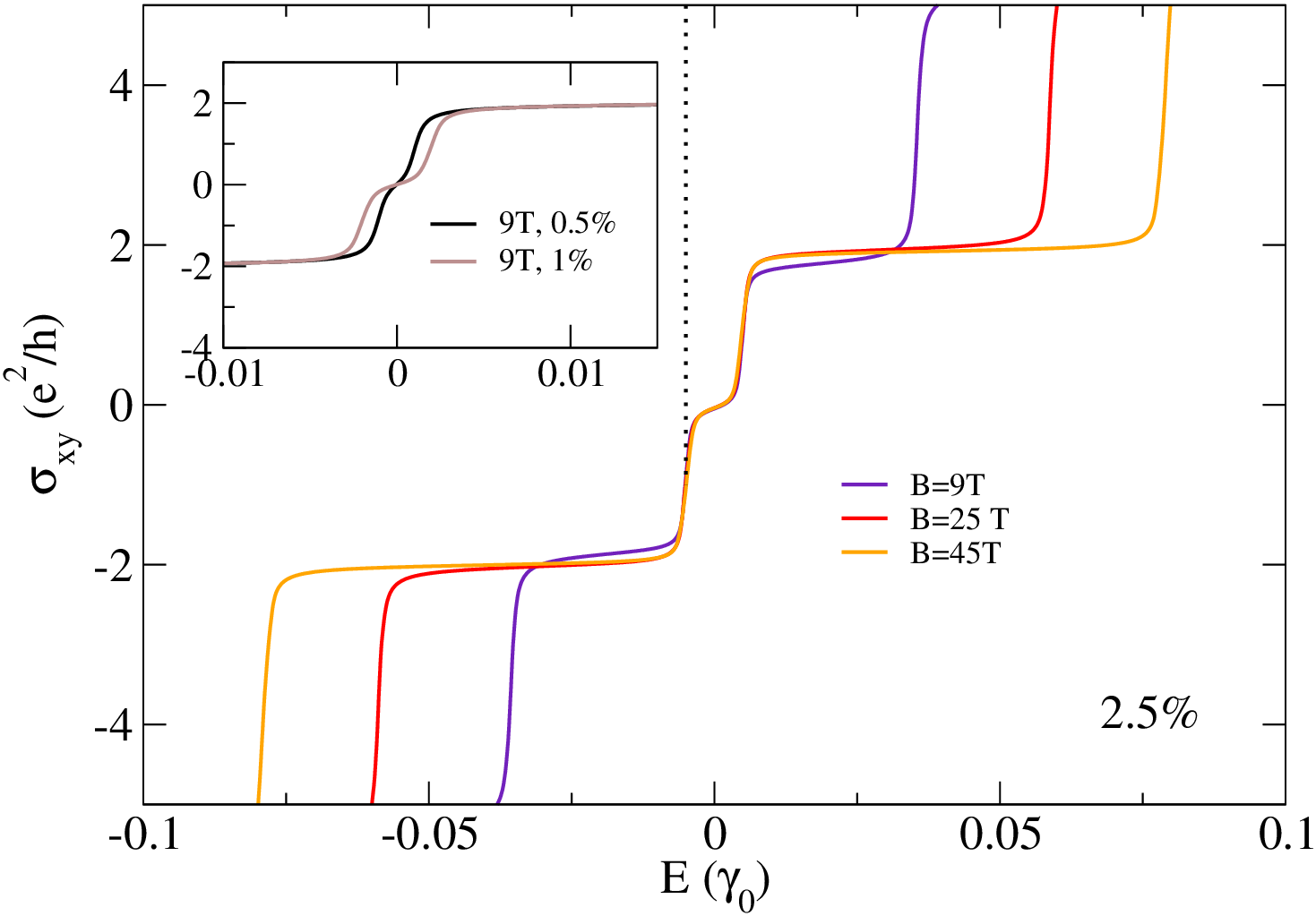}
\caption{(color online) Hall conductivity for p=2.5\% of AB-sublattice breaking defects (strenght $V_s=0.2\gamma_0$) distributed at random in space (see main text). Vertical dotted line indicates nominal gap of a correspondingly homogenous AB potential of strength $pV_{AB}=0.005\gamma_0$.}
\label{Fig3}
\end{center}
\end{figure}

As another challenging example, we present the analysis of a weakly correlated potential. We chose a globally sublattice-symmetry breaking potential that is locally disordered. In contrast to the above Anderson disorder model, which shifts all on-site energies at random, the present potential is only locally present, i.e. only for a fraction of sites that are selected at random. The disorder is chosen such that it breaks globally the AB-sublattice symmetry. For the results shown in Fig. \ref{Fig3} (main frame), we use a strength of $V_{AB}=0.2\gamma_0$ with $p=2.5\%$ of sites affected.
Besides ordinary conductance quantization at the values $\pm 2 e^2/h ,\pm 6 e^2/h,... $, an additional plateau is induced at zero energy with zero Hall conductivity.\footnote{The case of 45 T has been previously published in Ref. \cite{Ortmann:2013:QHE}.} The plateau is clearly visible for fields between 9 T and 45 T.
We find the width of the zero-energy plateau equal for all calculated magnetic fields. The corresponding energy value of the onset of the plateau at $\sigma=0$ is determined by the strength of the potential which is given as $pV_{AB}=0.005\gamma_0$ (indicated by dashed vertical line). The transition to higher plateaus (i.e. from $\pm2e^2/h$ to $\pm6e^2/h$) is rather influenced by the magnetic field-dependence of Landau level position.

It is interesting to study the emergence of such plateau when gradually increasing the concentration of impurities. Results are plotted for the case of 9T in Fig. \ref{Fig3} (inset). The plateau-length scaling with $p$ is evident from the figure. For the considered energy resolution in the inset ($\eta=0.0004$), a trace of plateau onset is still visible for the lowest percentages of the impurity potential down to 0.5\% (corresponding to $pV_{AB}=0.001\gamma_0$ or 2.7 meV).

\section{Conclusion} 
To conclude, an efficient real space algorithm to compute the Hall conductivity with the Kubo formula has been presented and validated on simple graphene-based systems. Such approach should become a useful computational tool to simulate QHE in very large size complex disordered materials such as polycrystalline graphene\cite{Cummings:PRB2014}, graphene subjected to weak van der Waals interaction such as by a boron-nitride substrate\cite{Martinez:2014PRB}, or other types of disordered two-dimensional materials. It should also allow to corroborate the formation of zero-energy plateaus of  $\sigma_{xy}$ driven by disorder-induced critical states, and disconnected from degeneracy lifting of Landau levels\cite{Leconte:2014:QHEEpoxy}, an issue of genuine fundamental interest in the context of topological interpretation of the quantized conductance.

\acknowledgements{F.O. would like to acknowledge the Deutsche Forschungsgemeinschaft for financial support within the Emmy Noether scheme (grant OR 349/1-1). ICN2 acknowledges support from the Spanish Ministry of Economy and Competitiveness under contract MAT2012-33911 the Severo Ochoa Program (MINECO, Grant SEV- 2013-0295).
We acknowledge PRACE for awarding us access to the Curie supercomputing center based in France. The support of Mikolaj Szydlarski from MDLS HPC Dev-Team, France to the technical work is gratefully acknowledged. }

\end{document}